\documentclass{caosp310}
\usepackage{graphicx}
\usepackage{natbib}
\usepackage{hyperref}
\usepackage{lscape}
\usepackage{cancel}
\usepackage{amsmath}

\articleNo{XXX}
\pubyear{XXXX}
\volume{XX}
\volnumber{X}
\firstpage{X}
\received{XXX XX, XXXX}
\accepted{XXX XX, XXXX}

\def\BibTeX{{\rm B\kern-.05em{\sc i\kern-.025em b}\kern-.08em
             T\kern-.1667em\lower.7ex\hbox{E}\kern-.125emX}}

\begin{document}

\hauthor{B. Dalla Barba et al.}
\htitle{Insights into jet–NLR energetics in PMN J0948+0022}

\title{Insights into jet–NLR energetics in \\ PMN J0948+0022}

\author{
        B. Dalla Barba \inst{1,2,3}\orcid{0009-0007-1729-2352}
       \and 
        L. Foschini\inst{2}\orcid{0000-0001-8678-0324}
      \and
        M. Berton\inst{3}\orcid{0000-0002-1058-9109}  
     \and 
     	A. L\"ahteenm\"aki\inst{4,5}\orcid{0000-0002-0393-0647}
      \and
      	M. Tornikoski\inst{4}\orcid{0000-0003-1249-6026}
      \and 
        E. Sani\inst{3}\orcid{0000-0002-3140-4070}
      \and 
        L. Crepaldi\inst{6,7}\orcid{0000-0001-7913-0577}
     \and 
        E. Congiu\inst{3}\orcid{0000-0002-8549-4083}
     \and 
     	G. Venturi\inst{8,9}\orcid{0000-0001-8349-3055}
     \and 
        W.J. Hon\inst{10}\orcid{0000-0001-8359-2328}
     \and 
        A. Vietri\inst{11,6}\orcid{0000-0003-4032-0853}
       }

\institute{\small{
        $^1$ Università degli studi dell’Insubria, Via Valleggio 11, Como 22100, Italy; \\
	$^2$ Osservatorio Astronomico di Brera, Istituto Nazionale di Astrofisica (INAF), Via E. Bianchi 46, Merate (LC) 23807, Italy; \\
	$^3$ European Southern Observatory (ESO), Alonso de Córdova 3107, Vitacura Santiago, Chile;\\
	$^4$ Aalto University Mets\"ahovi Radio Observatory, Mets\"ahovintie 114, FI-02540 Kylm\"al\"a, Finland;\\
	$^5$ Aalto University Department of Electronics and Nanoengineering, P.O. Box 15500, FI-00076 AALTO, Finland;\\
	$^6$ Dipartimento di Fisica e Astronomia "G. Galilei", Università degli studi di Padova, Vicolo dell'Osservatorio 3, Padova 35122, Italy; \\
	$^7$ Osservatorio Astronomico di Cagliari, Istituto Nazionale di Astrofisica (INAF), Via della Scienza 5, 09047 Selargius, Italy;\\
	$^8$ Scuola Normale Superiore, Piazza dei Cavalieri 7, Pisa 56126, Italy;\\
	$^9$ Osservatorio Astrofisico di Arcetri, Istituto Nazionale di Astrofisica (INAF), Largo E. Fermi 5, Firenze 50125, Italy;\\
	$^{10}$ School of Physics, University of Melbourne, Parkville, Victoria 3010, Australia; \\
	$^{11}$ Osservatorio Astronomico di Padova, Istituto Nazionale di Astrofisica (INAF), Vicolo dell'Osservatorio 5, 35122 Padova, Italy. \\
         \email{benedetta.dallabarba@inaf.it}}}

\date{month day, year}
\maketitle

\begin{abstract}
The analysis of the optical spectra of PMN J0948+0022 showed significant variations in the spectral lines that, when combined with the {\it Fermi} $\gamma-$ray light curve and radio observations reported by other authors, were interpreted as the result of interactions between the relativistic jet and the narrow-line region (NLR). In this work, we present order-of-magnitude calculations of the energetics associated with this proposed jet-NLR interaction. We demonstrate that the observed outflows are capable of absorbing a fraction of the jet energy and converting it into kinetic energy. This mechanism provides a natural explanation for the optical spectral variability recorded with the X-shooter and Multi-Unit Spectroscopic Explorer (MUSE) instruments. Our results support the scenario in which feedback from the relativistic jet can dynamically influence the circumnuclear gas, offering new insights into the coupling between jets and the NLR in $\gamma-$ray-emitting narrow-line Seyfert 1 galaxies.

\end{abstract}

\section{Introduction}
\label{intro}
Among active galactic nuclei (AGN), narrow-line Seyfert 1 galaxies (NLS1s) represent a peculiar class of objects. NLS1s are spectroscopically classified sources with relatively narrow H$\beta$, faint [O III]$\lambda$5007, and often strong Fe II multiplets \citep{1985ApJ...297..166O,1989ApJ...342..224G}. The importance of these sources lies in their nature as young or rejuvenated quasars \citep{2000MNRAS.314L..17M}, potentially representing the progenitors of more evolved quasars \citep{2016A&A...591A..88B,2017FrASS...4....8B,2025arXiv250903576B}. A small fraction of NLS1s show relativistic jet emission, which in some cases makes the sources detectable at $\gamma$-rays ($\gamma$-NLS1s). The first example of such a source is PMN J0948+0022 ($z=0.585$; \citealt{2002AJ....124.3042W,2003ApJ...584..147Z,2012ApJS..203...21A,2010ASPC..427..243F}). To date, about two dozen $\gamma$-NLS1s have been identified (e.g., \citealt{2022Univ....8..587F}), and their number is steadily increasing. 

In \citet{2025A&A...698A.320D} (hereafter referred to as DB25), we analyzed the optical spectral variations of PMN J0948+0022, focusing on the H$\beta$ and [O III]$\lambda\lambda4959,5007$ emission lines. We detected variability in the [O III]$\lambda5007$ core flux, in its outflow component, and in the outflow velocity. Based on intermediate-resolution ($R=\lambda/\Delta\lambda\sim6500-2500$) X-shooter (2017-12-17) and Multi-Unit Spectroscopic Explorer (MUSE) spectra (between 2022-11-24 and 2023-03-08, the analyzed spectrum combines data from all epochs), we proposed a reclassification of this source from a NLS1 to an intermediate Seyfert, characterized by a composite profile for the Balmer lines. We suggested that the X-shooter and MUSE H$\beta$ profiles are not due to a geometrical effect of partial obscuration, but rather to the interaction of the relativistic jet with the narrow-line region (NLR). This interaction was previously suggested by \citet{2019MNRAS.487..640D} from the study of the radio component of the jet. The authors found a jet-shape change at a distance of $\sim$100-400 pc from the center, compatible with our estimate of the distance of the outflow ($D_{out} \sim 130$-220 pc). From this, we proposed that the jet impacted the outflowing bubble, producing both the observed shape variation of the jet and the release of part of its energy into the surrounding NLR, which in turn generated the observed optical spectral variations.

In this work, we extend the study presented in DB25, focusing on the energetics of the jet-NLR interaction. These are back-of-the-envelope calculations, meant to provide an order-of-magnitude estimate of the process. The paper is organized as follows: Section~\ref{outflow} presents the physical conditions and energetics of the NLR; Section~\ref{jet} describes the energetics of the jet from $\gamma$-ray and radio data; Section~\ref{discussion} discusses the results and conclusions. Throughout the manuscript we assume a standard $\Lambda$CDM cosmology with H$_0$=73.3 km s$^{-1}$ Mpc$^{-1}$, $\Omega_{\rm m}$=0.3, and $\Omega_{\Lambda}$=0.7 \citep{2022ApJ...934L...7R}.

\section{NLR and outflow properties}
\label{outflow}
In general, from line ratios such as [O II] and [S II] we can estimate the electron density ($n_e$) and temperature ($T_e$) of the gas in the NLR. Given the X-shooter and MUSE spectral coverage, [S II] lines are not present in both PMN J0948+0022 spectra. For this reason, we assumed a typical $T_e \sim 10^4$ K \citep{2006agna.book.....O} and calculated the electron density from the [O II]$\lambda\lambda$3726,3729 lines using {\tt PyNeb} \citep{2013ascl.soft04021L,2015A&A...573A..42L}, a Python tool to compute emission line emissivities from recombination and collisionally excited lines (in our case we used only collisionally excited lines). The [O II] line fluxes from DB25 are reported in Table~\ref{tab:app}. The results, expressed as the median values obtained from N=5000 iterations with initial values inside the error bars of the observed fluxes, are: $n_{e,X}\sim170$ cm$^{-3}$ and $n_{e,M}\sim260$ cm$^{-3}$, for the X-shooter and MUSE cases, respectively.

From the calculated $n_e$ and the [O III]$\lambda$5007 luminosity of the outflow ($L_{5007}^{out}$), we can estimate the mass of the outflow ($M_{out}$). The $L_{5007}^{out}$ parameter was obtained from $L_{5007}^{out}=4\pi d_L^2 F_{5007}^{out}$, where $d_L$ is the luminosity distance and $F_{5007}^{out}$ is the observed flux of the outflowing component in [O III] (see again Table~\ref{tab:app}). $M_{out}$ is then calculated using the relation from \citet{2018MNRAS.477.5115K}:

\begin{equation}
M_{out} = 67.4 \times 10^7 
\left(\frac{L_{5007}^{out}}{10^{42} \, {\rm erg \, s}^{-1}}\right)\left(\frac{n_e}{100 \, {\rm cm}^{-3}}\right)^{-1} M_\odot
\end{equation}

\noindent where we re-scaled by a factor 10 the relation to use the [O III] luminosity instead of the H$\beta$ one, as suggested by \citet{2018MNRAS.477.5115K}. For the two cases, the results are: $M_{out,X} \sim 8.6 \cdot 10^7~M_\odot$ and $M_{out,M} \sim 3.9 \cdot 10^7~M_\odot$. 

Following the same reasoning presented in \citet{2005A&A...431..111B} and \citet{2023A&A...678A.127V}, the mass outflow rate through a spherical surface of radius \(r\) subtended by a solid angle \(\Omega\), with covering factor \(CF\), is \(\dot{M}_{out} = \Omega r^2 \rho v_{out}\cdot CF\). If the outflowing mass \(M_{out}\) is contained in a thin shell of radial thickness \(R_{out}\), its volume is approximately \(V \simeq \Omega r^2 R_{out}\) and therefore:

\begin{equation}
\begin{split}
  \dot{M}_{out} 	& \sim  \Omega r^2 \frac{M_{out}}{V} v_{out}\cdot CF \sim \cancel{\Omega} \frac{M_{out}}{\cancel{\Omega r^2} R_{out}} \cancel{r^2}  v_{out} \cdot CF \sim \\
  			& \sim \frac{M_{out}}{R_{out}} v_{out} \cdot CF\sim \frac{M_{out}}{D_{out}\cdot\cancel{0.1}} v_{out} \cdot \cancel{0.1}
\end{split}
\end{equation}

\noindent assuming roughly constant density and velocity across the shell, we have adopted a reference covering factor $CF \sim 0.1$ and an outflow distance from the center of $D_{out} \sim R_{out}/0.1 \sim 173\ \rm pc$ (see the mean value reported in Table~\ref{tab:app}). In the $D_{out}$ expression, we have further assumed a conical shape for the jet. Finally, the kinetic power of the outflow is:

\begin{equation}
  \dot{E}_{kin} = \frac{1}{2}\dot{M}_{out} v_{out}^2=P_{out}
\end{equation}

The resulting estimates for the two cases are $P_{out,X} \sim 8.8\cdot 10^{42}\ \mathrm{erg\ s^{-1}}$ and $P_{out,M} \sim 14\cdot 10^{42}\ \mathrm{erg\ s^{-1}}$.\\
If we instead assume electron temperatures in the range $T_e = (5-20)\cdot10^4$ K \citep{2006agna.book.....O}, the resulting densities are $n_{e,X} \sim 120-210\ \mathrm{cm^{-3}}$ and $n_{e,M} \sim 190-320\ \mathrm{cm^{-3}}$. The corresponding outflow powers are $P_{out,X} \sim (7.0-12)\cdot10^{42}\ \mathrm{erg\ s^{-1}}$ and $P_{out,M} \sim (12-19)\cdot10^{42}\ \mathrm{erg\ s^{-1}}$. In all cases, these variations do not affect the conclusions.

\section{Jet power}
\label{jet}
\subsection{$\gamma$-ray component}
\label{gamma}
We calculated the radiative power of the jet ($P_{\gamma, \rm rad}$) from {\it Fermi} data obtained in epochs nearly simultaneous with the X-shooter and MUSE observations. For X-shooter, we used the photon flux from 2017-12-30, while for MUSE we calculated the mean value of the {\it Fermi} datapoints over the interval 2022-12-04 to 2023-02-02. We multiplied the photon flux ($F_\gamma$, reported in Table~\ref{tab:app}) by the photon mean energy, using 1 GeV as reference for the conversion, and then obtained the flux in erg s$^{-1}$ cm$^{-2}$. We calculated the corresponding luminosity ($L_{\gamma}$) using:

\begin{equation}
	L_\gamma=4 \pi d_L^2 \frac{F_\gamma}{(1+z)^{1-\alpha_\gamma}}
\end{equation}

\noindent where $\alpha_\gamma$ is the spectral index in the $\gamma$-rays obtained from the {\it Fermi} light-curve repository (see Table~\ref{tab:app}). With the luminosity, we then calculated $P_{\gamma,rad}$ using the equation from \citet{2003ApJ...593..667M}:

\begin{equation} 
P_{\gamma,rad}=\Gamma^2 \frac{L_\gamma}{\delta^4} 
\end{equation}

\noindent where $\Gamma$ is the bulk Lorentz factor and $\delta$ the corresponding Doppler factor. As reported in Table~\ref{tab:app}, we can assume two possible values for these quantities, which lead to four different results. Here we report only the two mean values: $P_{\gamma,\mathrm{rad},X} \sim 2.5 \times 10^{44}$ erg s$^{-1}$ and $P_{\gamma,\mathrm{rad},M} \sim 2.4 \times 10^{44}$ erg s$^{-1}$. Finally, we assume that the kinetic power is ten times the radiative one, see Table~\ref{tab:power} for $P_{\gamma,kin}$.

\subsection{Radio component}
\label{radio}
For the radio component, we can use the relation presented in \citet{2024Univ...10..156F} to obtain kinetic ($P_{radio,kin}$) power of the jet:

\begin{equation}
	P_{radio,kin}=3.9\times10^{44}\left(\frac{S_\nu d_L^2}{1+z}\right)^{\frac{12}{17}}~{\rm erg~s}^{-1}
\end{equation}

\noindent where $S_{\nu}$ is the radio flux density in Jy (see Table~\ref{tab:app}). The radio data were taken from the monitoring of the source performed by the Metsähovi Radio Observatory at 37 GHz. For the X-shooter epoch we used the flux averaged between 2017-12-06 and 2017-12-22, and for the MUSE epochs the flux averaged between 2022-12-02 and 2023-03-06. The results are: $P_{radio,kin,X}\sim 4.8\cdot 10^{44}$ erg s$^{-1}$ and $P_{radio,kin,M}\sim 7.3\cdot 10^{44}$ erg s$^{-1}$.

\section{Discussion and conclusions}
\label{discussion}
From the values reported in reported Table~\ref{tab:power}, we can compare the kinetic powers of the outflow with those of the jet in its $\gamma-$ray and radio components. Specifically, we find that the outflow kinetic power constitutes a small but non-negligible fraction of the jet power in both epochs. The respective ratios, listed in Table~\ref{tab:power}, indicate that approximately 0.35\%-1.9\% of the jet power has been deposited into the outflow in both epochs. This further supports the scenario proposed in DB25, where we suggested that the observed variations in the outflow properties (in terms of flux and velocity) could result from the interaction between the jet and the NLR. In turn, these variations in the outflow kinetic power would likely be driven by changes in the jet state and energetics.

These results support a scenario in which the relativistic jet deposits part of its energy into the surrounding environment -- specifically, the NLR -- producing the observed variations in the optical spectra, particularly in the [O III]$\lambda5007$ outflow properties. This hypothesis is also supported by the radio study of \citealt{2019MNRAS.487..640D}. Our optical spectroscopic analysis (DB25) provides a crucial link between these radio observations and the optical data, linking changes in jet morphology with the variability observed in both the optical band and the $\gamma$-rays. The order-of-magnitude calculations presented here provide further evidence that, from an energetic perspective, the jet-NLR interaction can account for the observed optical spectral variations. Together, these findings reinforce the notion that the interplay between the relativistic jet and the circumnuclear gas significantly influences both the kinematics and energetics of the NLR in $\gamma-$NLS1 galaxies. Similar results have been reported by other authors, who observed variability in spectral features associated with jet activity. Examples include variable Mg II lines (e.g., \citealt{2013ApJ...763L..36L,2018A&A...614A.148B,2020MNRAS.493.5773Y}) and changes in the broad H$\beta$ component \citep{2023A&A...672L..14H}.

In conclusion, this analysis, combined with the results reported in DB25, underscores the importance of a multi-wavelength approach to the study of AGN, particularly for jetted sources. The combination of multi-epoch optical spectroscopy, radio observations, and $\gamma-$ray monitoring is essential for a better understanding of the mechanisms driving the interplay between jet emission and the AGN environment.

\begin{table}[htbp]
\caption{Summary of the kinetic powers and of the ratios between the outflow power and the $\gamma-$ray and radio jet powers ($R_{out,\gamma}$ and $R_{out,radio}$, respectively).}
\label{tab:power}
\begin{center}
\renewcommand{\arraystretch}{1.25}
\begin{tabular}{l | lll | ll}
\hline\hline
	{\bf Epoch} 		& {\bf $P_{out}$} 		& {\bf $P_{\gamma,kin}$} 			& {\bf $P_{radio,kin}$}	& {\bf $R_{out,\gamma}$}	& {\bf $R_{out,radio}$} \\
	{\bf}				& {\bf [erg s$^{-1}$]}		& {\bf [erg s$^{-1}$]}				& {\bf [erg s$^{-1}$]} 		& {\bf}				& {\bf} \\
	\hline
	X-shooter			& $8.8\cdot 10^{42}$		& $ 2.5 \cdot 10^{45}$			& $4.8\cdot 10^{44}$		& 0.35\%				& 0.55\%	\\
	MUSE			& $14\cdot 10^{42}$		& $ 2.4 \cdot 10^{45}$			& $7.3\cdot 10^{44}$		& 0.58\%				& 1.9\%	\\
\hline\hline
\end{tabular}
\renewcommand{\arraystretch}{1}
\end{center}
\end{table}

\acknowledgements
B.D.B. thank the organizers of the 15th Serbian Conference on Spectral Line Shapes in Astrophysics for the contributed talk. G.V. acknowledges support from the European Union (ERC, WINGS, 101040227). This publication makes use of data obtained at Mets\"ahovi Radio Observatory, operated by Aalto University in Finland.

\bibliography{biblio.bib}

\begin{landscape}
\appendix{Useful quantities}
\begin{table}[htbp]
\caption{List of the quantities involved in the calculations. References: {\it F}LCR) {\it Fermi} light-curve repository \citep{2023ApJS..265...31A}, F12) \citet{2012A&A...548A.106F}, {\it Mets}) Mets\"ahovi Radio Observatory data.}
\label{tab:app}
\begin{center}
\footnotesize
\renewcommand{\arraystretch}{1.25}
\begin{tabular}{lllll}
\hline\hline
	{\bf Quantity} 								& {\bf Symbol(s)} 		& {\bf Value(s)} 					& {\bf Units} 			& {\bf Reference} \\
\hline
	Luminosity distance							&	$d_L$			&	$3.2$					&	Gpc				&	Redshift	\\
	Outflow distance							&	$D_{out}$			&	130/220 $\rightarrow$ 170		&	pc				&	DB25	\\
	X-shooter [O II]$\lambda$3726 flux				&	$F_{3726,X}$ 		&	$(3.0 \pm 0.6)\times 10^{-17}$	&	erg s$^{-1}$ cm$^2$	&	DB25 \\
	X-shooter [O II]$\lambda$3729 flux				&	$F_{3729,X}$ 		&	$(4.3 \pm 0.4)\times 10^{-17}$	&	erg s$^{-1}$ cm$^2$	&	DB25 \\
	MUSE [O II]$\lambda$3726 flux				&	$F_{3726,M}$ 		&	$(8.4 \pm 0.6)\times 10^{-17}$	&	erg s$^{-1}$ cm$^2$	&	DB25 \\
	MUSE [O II]$\lambda$3729 flux				&	$F_{3729,M}$ 		&	$(9.9 \pm 0.6)\times 10^{-17}$	&	erg s$^{-1}$ cm$^2$	&	DB25 \\
	{\it Fermi} flux in the X-shooter epoch			&	$F_{\gamma,X}$	&	$4.0 \times 10^{-8}$			&	ph s$^{-1}$ cm$^2$	&	{\it F}LCR \\
	{\it Fermi} flux in the MUSE epochs				&	$F_{\gamma,M}$	&	$3.8 \times 10^{-8}$			&	ph s$^{-1}$ cm$^2$	&	{\it F}LCR \\
	X-shooter [O III]$\lambda$5007	 outflow luminosity	&	$L_{5007,X}^{out}$ 	&	$ 2.1 \times 10^{41}$		&	erg s$^{-1}$		&	DB25 \\
	MUSE [O III]$\lambda$5007 outflow luminosity		&	$L_{5007,M}^{out}$ 	&	$ 1.5 \times 10^{41}$		&	erg s$^{-1}$		&	DB25 \\
	Radio flux density close to the X-shooter epoch 	&	$S_{\nu,X}$		&	0.20						&	Jy				&	{\it Mets} \\
	Radio flux density close to the MUSE epochs 		&	$S_{\nu,M}$		&	0.37						&	Jy				&	{\it Mets} \\
	X-shooter [O III]$\lambda$5007 outflow velocity 	& 	$v_{out,X}$		&	380						&	km s$^{-1}$		& 	DB25 \\
	MUSE [O III]$\lambda$5007 outflow velocity		&	$v_{out,M}$		&	580						&	km s$^{-1}$		& 	DB25 \\
	{\it Fermi} spectral index						&	$\alpha_{\gamma}$	&	2.5						&	-				&	{\it F}LCR \\
	Bulk Lorentz factor (two cases)					&	$\Gamma_{1,2}$	&	11/16					&	-				&	F12 \\
	Doppler factor	(two cases)					&	$\delta_{1,2}$		&	17/19					&	-				&	F12 \\
\hline\hline
\end{tabular}
\renewcommand{\arraystretch}{1}
\end{center}
\end{table}
\end{landscape}

\end{document}